\documentclass[twocolumn]{jpsj2} 
\usepackage{bm}

\title{ Spin Fluctuations and Unconventional Superconductivity in the Fe-based Oxypnictide Superconductor LaFeAsO$_{0.7}$ probed by $^{57}$Fe-NMR}

\author{
Nobuyuki Terasaki$^{1}$, Hidekazu Mukuda$^{1}$\thanks{E-mail address: mukuda@mp.es.osaka-u.ac.jp},  Mitsuharu Yashima$^{1}$, Yoshio Kitaoka$^{1}$,\\  Kiichi Miyazawa$^{2}$, Parasharam Shirage$^{2}$, Hijiri Kito$^{2}$, Hiroshi Eisaki$^{2}$, Akira Iyo$^{2}$ 
}

\inst{
$^{1}$Graduate School of Engineering Science, Osaka University, Toyonaka, Osaka 560-8531 \\
$^{2}$National Institute of Advanced Industrial Science and Technology (AIST), Umezono, Tsukuba 305-8568 \\
}

\abst{
We report $^{57}$Fe-NMR studies on the oxygen-deficient iron (Fe)-based oxypnictide superconductor LaFeAsO$_{0.7}$ ($T_{c}=$ 28 K) enriched by $^{57}$Fe isotope. In the superconducting state, the spin component of $^{57}$Fe-Knight shift $^{57}K$ decreases almost to zero at low temperatures and the nuclear spin-lattice relaxation rate $^{57}(1/T_{1})$ exhibits a $T^{3}$-like dependence without the coherence peak just below $T_{c}$, which give firm evidence of the unconventional superconducting state formed by spin-singlet Cooper pairing.  All these events below $T_c$ are consistently argued in terms of the extended s$_{\pm}$-wave pairing with a sign reversal of the order parameter among Fermi surfaces. In the normal state, we found the remarkable decrease of $1/T_1T$ upon cooling for both the Fe and As sites, which originates from the decrease of low-energy spectral weight of spin fluctuations over whole ${\bm q}$ space upon cooling below room temperature. Such behavior has never been observed for other strongly correlated superconductors where an antiferromagnetic interaction plays a vital role in mediating the Cooper pairing. 
} 

\kword{superconductivity, iron-based oxypnictide, LaFeAsO, NMR}

\begin{document}

\maketitle

\date{\today}


After the discovery of iron (Fe)-based layered oxypnictide superconductor LaFeAs(O$_{1-x}$F$_{x}$),\cite{Kamihara} the replacement of La site by other rare earth elements significantly enhances a superconducting transition temperature $T_{c}$ up to more than 50 K.\cite{Ren1,Kito,Ren2}  A new route has opened up to deepen understanding of high-$T_c$ superconductivity (SC) phenomena. 
Mother material LaFeAsO is a semimetal with a stripe antiferromagnetic (AFM) order with ${\bf Q}=(0,\pi)$ or $(\pi,0)$.  
The crystal structure contains alternate stacking of LaO and FeAs layers along the c-axis. In this structure, the Fe atoms of the FeAs layer are located in a four-fold coordination forming a FeAs$_{4}$-tetrahedron. 
Substitution of oxygen site with fluorine atoms and/or oxygen deficiency at LaO layer causes an exotic SC with a highest $T_c$ besides copper-oxides superconductors.\cite{Kamihara,Ren1,Kito,Ren2} 
Remarkably, Lee et al. have found that $T_{c}$ increases up to a highest value of $T_c$=54 K when FeAs$_{4}$ tetrahedron is transformed toward a regular one.\cite{C.H.Lee}  Relevant with this fact, we reported that the nuclear quadrupole frequency at the As site relates to $T_c$ for various FeAs-based oxypnictide superconductors, unraveling that $T_c$ is sensitive to the local configuration of FeAs$_{4}$ tetrahedron.\cite{Mukuda}  
The present experimental facts suggest that systematic understanding of local electronic state at the Fe site is quite important to elucidate the origin of SC in the iron-based compounds.

In this letter, we report $^{57}$Fe-NMR studies on the superconducting and normal-state properties of LaFeAsO$_{0.7}$ with $T_c=$ 28 K enriched by $^{57}$Fe isotope. It is reinforced that the $^{57}$Fe-NMR-$(1/T_1)$ exhibits a $T^{3}$-like dependence without the coherence peak just below $T_{c}$ and the  spin component of $^{57}$Fe-Knight shift decreases almost to zero at low temperatures. These results pointing to the unconventional SC with spin-singlet Cooper pairings are consistent with the theoretical model which proposes that an extended s$_{\pm}$-wave pairing realizes with a sign reversal of the order parameter among Fermi surfaces. 
In the normal state, we discuss the characteristics of spin fluctuations in this compound by comparing $^{57}$Fe-$(1/T_{1})$ with the $^{75}$As-$(1/T_{1})$. 
 

The polycrystalline sample of $^{57}$Fe-enriched LaFeAsO$_{0.7}$ was synthesized via the high-pressure synthesis technique, as described in the elsewhere.\cite{Kito} 
In particular, starting materials of LaAs, $^{57}$Fe, $^{57}$Fe$_{2}$O$_{3}$ enriched by the $^{57}$Fe isotope ($^{57}$Fe : nuclear spin $I$ = 1/2, $^{57}\gamma_n/2\pi= 1.3757$ MHz/T) were mixed with nominal composition of LaFeAsO$_{0.7}$. 
The powder X-ray diffraction measurement indicates that the sample composes of almost a single phase. The superconducting transition temperature $T_c$ was determined to be 28 K by distinct decrease in the dc susceptibility due to the onset of SC diamagnetism. Although the real content of oxygen is unclear, the lattice parameters $a$ = 4.0226\r{A} and $c$ = 8.7065\r{A}  are almost the same as the LaFeAsO$_{0.6}$
 with $T_c$=28 K in the previous work\cite{Mukuda}, which indicates that both samples are compatible in their physical properties. 
The $^{57}$Fe-NMR measurements were performed in the moderately crushed powder sample, which was oriented along the direction including the $ab$ plane.

\begin{figure}[tbp]
\begin{center}
\includegraphics[width=0.9\linewidth]{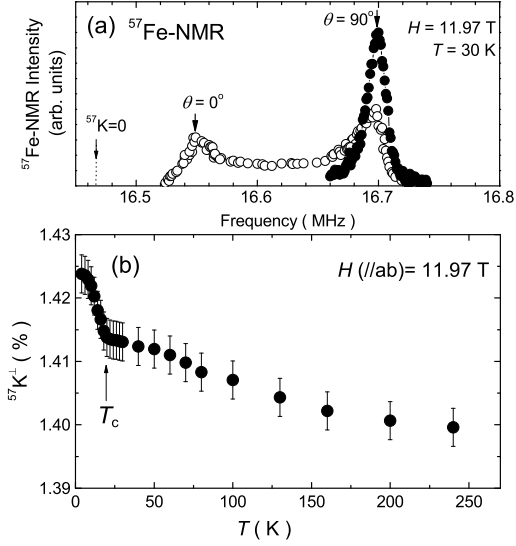}
\end{center}
\caption[]{
(a) $^{57}$Fe-NMR spectra at $H$=11.97 T and 30 K for LaFeAsO$_{0.7}$ in the field parallel($\bullet$) and perpendicular($\circ$) to the orientation direction. 
(b) The $T$ dependence of $^{57}K^\perp$ at $H$=11.97 T ($T_{c}(H)\sim$ 20 K). It is noteworthy that the $T$-dependence of $^{57}K^\perp$ is opposite to those of $^{75}$As and $^{19}$F  sites \cite{Grafe,Ahilan,ImaiJPSJ}, indicating that the hyperfine-coupling constant is negative at Fe site, originating from the inner core-polarization.
}
\label{spectrum}
\end{figure}

Figure \ref{spectrum}(a) shows $^{57}$Fe-NMR spectra obtained by sweeping frequency ($f$) at magnetic field $H=$ 11.97 T at 30 K. 
When the field is applied parallel to the orientation direction, we observed a single peak, the linewidth of which is as narrow as $\sim$20 kHz at 11.97 T, demonstrating that the FeAs layers of this sample are homogeneous irrespective of the oxygen deficiency in the LaO layer.  
In the field perpendicular to the orientation direction, two horn peaks are observed, which arises from crystals with $\theta=90^\circ$ and $0^\circ$, where $\theta$ is the angle between field and c-axis of the crystal. 
Knight shifts of $^{57}$Fe, that is determined by the shift from $f_{0}=~^{57}\gamma_n H$ ($^{57}K=0$), are $^{57}K^\perp \sim 1.413$\% and $^{57}K^\parallel \sim 0.50$\% at 30 K for $\theta=90^\circ$ and $0^\circ$, respectively. 

\begin{figure}[tbp]
\begin{center}
\includegraphics[width=0.9\linewidth]{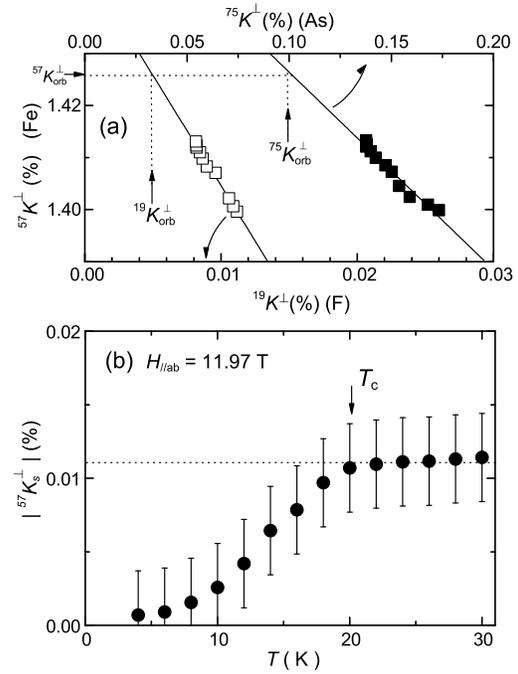}
\end{center}
\caption[]{(a) $^{57}K^\perp$(Fe) plotted against $^{75}K^\perp$(As) and $^{19}K^\perp$(F) reported by Imai {\it et al.} \cite{ImaiJPSJ}, the slopes of which gives the ratio of $^{57}A_{\rm hf}^\perp/^{75}A_{\rm hf}^\perp\simeq -0.38$ and $^{57}A_{\rm hf}^\perp/^{19}A_{\rm hf}^\perp\simeq -4.2$. $^{57}K_{\rm orb}^\perp\sim 1.425$\% is deduced using the orbital shift of $^{19}$F and $^{75}$As sites\cite{ImaiJPSJ}.
(b) $|^{57}K_{\rm s}^\perp|$ evaluated from $|^{57}K^\perp-^{57}K_{\rm orb}^\perp|$ decreases to zero in the SC state, which gives firm evidence for spin-singlet Cooper pairing state. 
}
\label{Knightshift}
\end{figure}

The Knight shift comprises generally of the $T$-independent orbital contribution and $T$-dependent spin contribution, denoted as $^{57}K_{\rm orb}$ and $^{57}K_{\rm s}$, respectively. 
Here, $^{57}K=~^{57}K_{\rm orb}+~^{57}K_{\rm s}$ with $^{57}K_{\rm s}=~^{57}A_{\rm hf}\chi_{\rm s}$, where $^{57}A_{\rm hf}$ is the hyperfine coupling constant and $\chi_{\rm s}$ is the uniform spin susceptibility.  
Figure \ref{spectrum}(b) shows the $T$ dependence of $^{57}K^\perp$ in the field parallel to the $ab$-plane. It is noteworthy that the $T$ dependence of $^{57}K^\perp$ is opposite to those of $^{75}$As and $^{19}$F sites\cite{Grafe,Ahilan,ImaiJPSJ}, indicating that $^{57}A_{\rm hf}^\perp$ is negative at Fe site, originating from the inner core-polarization.  
In this compound, $^{57}A_{\rm hf}^\perp$ at the Fe site is given by $A+4B$, where $A$ is the on-site negative term dominated by the inner core-polarization, and $B$ is the transferred positive one from the neighbour Fe site through the direct Fe-Fe and/or the indirect Fe-As-Fe bondings. 
Here, the $B$ composes of $B_1$ and $B_2$ derived from the first and second nearest neighbour Fe sites, respectively.
In contrast, the $^{75}A_{\rm hf}^\perp$ at the As site is given by $4C$, where $C$ is the positive term derived from the transferred field($C_{\rm tr}$) and diagonal pseudo-dipole field($C_{\rm dip}$) induced by one neighboring Fe site. Note that $C$ is deduced to be $\sim$ 6.2~kOe/$\mu_{\rm B}$.\cite{Grafe}  
Figure \ref{Knightshift}(a) shows $^{57}K^\perp(T)$ plotted against those of $^{75}$As and $^{19}$F sites reported by Imai {\it et al.} \cite{ImaiJPSJ}, as the $T$ implicit parameter. Their slopes indicate a ratio of hyperfine coupling constants, $^{57}A_{\rm hf}^\perp/^{75}A_{\rm hf}^\perp \simeq -0.38$ and $^{57}A_{\rm hf}^\perp/^{19}A_{\rm hf}^\perp\simeq -4.2$ for ${\bm q}=0$ component. 
Assuming $A \sim -139$ kOe/$\mu_{\rm B}$ that is known as the theoretical value of core polarization field for Fe$^{2+}$ ion\cite{FreemanWatson}, $B\sim$ 32.4~kOe/$\mu_{\rm B}$ is deduced from $^{57}A_{\rm hf}^\perp/^{75}A_{\rm hf}^\perp=(A+4B)/(4C)=-0.38$.
As shown in Fig. \ref{Knightshift}(a), $^{57}K_{\rm orb}^\perp$ was estimated to be $1.425$\% by assuming the orbital shifts of $^{19}$F and $^{75}$As sites\cite{ImaiJPSJ}, which enables us to evaluate $|^{57}K_{\rm s}^\perp|$ by $|^{57}K^\perp-^{57}K_{\rm orb}^\perp|$. 
Remarkably, the $|^{57}K_{\rm s}^\perp|$ decreases to almost zero well below $T_c$, as shown in Fig. \ref{Knightshift}(b), suggesting possible existence of an isotropic gap in very low temperature regime.  This result gives firm evidence of spin-singlet Cooper pairing through the first $^{57}$Fe-Knight shift measurement. 

The nuclear spin-lattice relaxation rate $^{57}(1/T_1)$ of $^{57}$Fe was determined from the recovery curve of $^{57}$Fe nuclear magnetization, which is expressed by a simple exponential function as $m(t)\equiv(M(\infty)-M(t))/M(\infty)=\exp(-t/T_{1})$. Here $M(\infty)$ and $M(t)$ are the respective nuclear magnetizations for the thermal equilibrium condition and at a time $t$ after the saturation pulse. Note that $^{57}T_1$ was uniquely determined from a single exponential function of $m(t)$ in the whole $T$ range, suggesting that the normal-state properties are homogeneous and the presence of vortex cores in the SC mixed state under $H$ does not prevent from probing quasiparticle excitations inherent to the novel SC state in LaFeAsO$_{0.7}$. 
We measured the $^{57}(1/T_1)$ in the field along the $ab$-plane, since it was the same with that along  the $c$-axis at 30K. 

\begin{figure}[tbp]
\begin{center}
\includegraphics[width=0.9\linewidth]{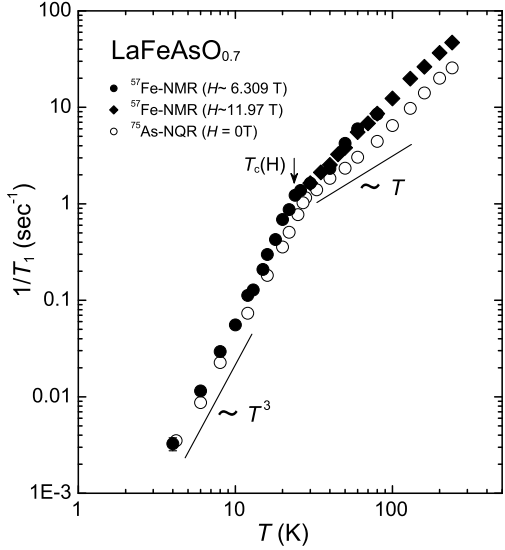}
\end{center}
\caption[]{ $T$ dependence of $^{57}$Fe-NMR $1/T_{1}$ at $H$= 6.309T and $H=$ 11.97 T, along with the $^{75}$As-NQR $1/T_{1}$ for LaFeAsO$_{0.6}$($T_{c}$ = 28 K).\cite{Mukuda}
In the SC state, $^{57}$Fe-$1/T_{1}$ follows a $T^3$-like dependence upon cooling without the coherence peak just below $T_{c}(H)$ = 24 K, which resemble the result by $^{75}$As-NQR.
}
\label{T1}
\end{figure}

Figure~\ref{T1} shows the $T$ dependences of $^{57}(1/T_1)$ at $H =$ 6.309 T and 11.97 T in the $T$ range of 4$\sim$80 K and 30$\sim$240 K, respectively. In the SC state, the $^{57}$Fe-NMR $(1/T_{1})$ exhibits a $T^3$-like dependence without the coherence peak just below $T_{\rm c}(H)=$ 24 K at $H$ = 6.309 T, which resembles the previous $^{75}$As-NQR $(1/T_{1})$ in LaFeAsO$_{0.6}$.\cite{Mukuda}  It is noteworthy that the residual density of states (RDOS) well below $T_{\rm c}$ was not observed at the Fermi level even for $T/T_{\rm c}=0.17$, although RDOS used to be observed in the unconventional superconductors with the line-node gap, easily induced by impurity scatterings especially in a unitality limit. 
These features of $1/T_1$ were commonly reported among other Fe-based superconductors \cite{Nakai,Matano,Grafe,Mukuda}. In contrast,  a fully gapped SC state was revealed  by the experiments such as ARPES\cite{ARPES} and magnetic penetration depth.\cite{Penetrationdepth} 
In order to reconcile these issues, the theoretical groups have carried out the calculation of $1/T_1$ on the basis of a nodeless extended $s_\pm$-wave pairing model with a sign reversal of the order parameter between the hole and electron Fermi surfaces\cite{Mazin,Kuroki}.  In the framework of either a two-band model where the unitaly scattering due to impurities is assumed \cite{NMRtheory} or a five-band model in a rather clean limit\cite{NMRtheory2}, the experiments are well reproduced by such the calculations. As a matter of fact, the results of $(1/T_1)$s for $^{57}$Fe and $^{75}$As in the SC state are consistent with the latter model. This may be because the intrinsic behavior of $1/T_1$ is measured on the highly homogeneous sample, that is guaranteed by the very sharp NMR linewidth with as small as 20 kHz even at 12 T.
In this context, our results is consistently argued in terms of the extended s$_{\pm}$-wave pairing with a sign reversal of the order parameter among Fermi surfaces.  

\begin{figure}[tbp]
\begin{center}
\includegraphics[width=0.9\linewidth]{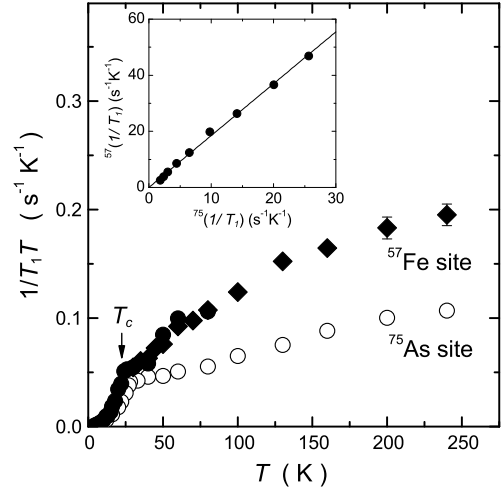}
\end{center}
\caption[]{ $T$ dependence of $^{57}$Fe-$1/T_{1}T$ at $H$ = 6.309 T ($\bullet$) and $H=$ 11.97 T(closed diamond), together with the $^{75}$As-$1/T_{1}T$ in LaFeAsO$_{0.6}$($\circ$),\cite{Mukuda} respectively.
The inset shows the plot of $^{57}(1/T_{1}T)$ vs $^{75}(1/T_{1}T)$  as the implicit parameter of $T$ between 30 K and 240 K.
}
\label{T1T}
\end{figure}

Next we deal with the normal-state properties probed by the $(1/T_1)$s of both $^{57}$Fe and $^{75}$As.
As shown in Fig. \ref{T1T}, $^{57}(1/T_{1}T)$ in the normal state gradually decreases upon cooling down to $T_{c}$, resembling that of $^{75}(1/T_{1}T)$ at As site. 
Actually, $^{57}(1/T_{1}T)$ at Fe site is well scaled to $^{75}(1/T_{1}T)$ at As site \cite{Mukuda} down to 60 K with a ratio of  $^{57}(1/T_{1}T)/^{75}(1/T_{1}T)\simeq 1.85$, as shown in the inset, whereas it deviates slightly from this linear relation in the range of $T=$30 K$-$60 K. 
The respective $1/T_{1}$s of $^{57}$Fe and $^{75}$As are expressed as,
\begin{equation}
^{57,75}\left(\frac{1}{T_{1}}\right)\sim \sum_{\bm q}(^{57,75}\gamma_{n})^{2}|^{57,75}A_{\rm hf}({\bm q})|^{2}\frac{\chi ''({\bm q},\omega_{0})}{\omega_{0}},
\end{equation}
\begin{equation}
\begin{array}{ll}
^{57}A_{\rm hf}(\bm q)=A+2B_1[\cos(q_x a)+\cos(q_y a)]\\
\ \ \ \ \ \ \ \ \ \ \ \ \ \ \ \ \ \ \ +4B_2[\cos(q_x a)\cos(q_y a)],\\
^{75}A_{\rm hf}(\bm q)=4C_{\rm tr}[\cos(q_x a/2)\cos(q_y a/2)]+ 4C_{\rm dip}^\perp({\bm q}),\\
\end{array}
\end{equation}
where $A_{\rm hf}({\bm q})$ is the wave-vector (${\bm q}$)-dependent hyperfine coupling constant, $\chi''({\bm q},\omega_{0})$ is the imaginary part of the dynamical spin susceptibility, $a$ is a distance between Fe atoms, and $\omega_{0}$ is the NMR/NQR frequency. 
An anisotropy of $^{57}(1/T_1)$ between the $ab$-plane and the $c$-axis is negligibly small for Fe site, which ensures that $\chi''({\bm q},\omega_{0})$ and the hyperfine fields ($A$ and $B$) are  isotropic. 
Although the anisotropy of $^{75}(1/T_1)$ was observed in other $^{57}$As-NMR measurement \cite{Grafe}, it is predominantly derived from the anisotropy of $C_{\rm dip}^\perp({\bm q})$, which is the inplane component of off-diagonal pseudo-dipole hyperfine field arising from the $\chi''({\bm q},\omega_{0})$ along the $c$-axis. 

Now let us consider three simple cases on the basis of the possible evaluation of hyperfine field.
(i) First, we assume a case that spin fluctuations only around ${\bm q}=0$ are predominant. In this case, we would expect a ratio of $^{57}(1/T_{1}T)/^{75}(1/T_{1}T)\sim$0.005 by using $(^{57}A_{\rm hf}^\perp/^{75}A_{\rm hf}^\perp)^2=(-0.38)^2$. However, it is three orders of magnitude smaller than the experimental value of 1.85 at $T=$ 60$\sim$250 K, suggesting that the ferromagnetic spin-fluctuation modes around ${\bm q}=0$ are not predominant in this compound. 
(ii) Second, we assume the case that the spin fluctuations only around ${\bm Q}=(0,\pi/a)$ or $(\pi/a,0)$ are predominant. 
It is characteristic that the $|^{57}A_{\rm hf}({\bm q})|$ at the Fe site becomes large around the boundaries of the first Brillouin zone. 
For example, $|^{57}A_{\rm hf}(0, \pi/a)|=|A-4B_2|$ is roughly evaluated to be $\sim$204 kOe/$\mu_{\rm B}$ if $B_1$ and $B_2$ were comparable, which are one order of magnitude larger than that of As site. 
Even though the off-diagonal term $4C_{\rm dip}^\perp({\bm q})$ in eq.(2) is not clear for the paramagnetic state of LaFeAsO$_{0.7}$, it is tentatively assumed to be 4.3 kOe/$\mu_{\rm B}$ obtained for the stripe AFM ordered phase in the orthorhombic BaFe$_2$As$_2$\cite{Kitagawa}. 
Then we obtain $^{57}(1/T_{1}T)/^{75}(1/T_{1}T)\sim 5$, which is the same order of magnitude to the experimental value. 
(iii) Third, we assume the case that the spin fluctuations only around ${\bm Q}=(\pi/a,\pi/a)$ are predominant. In this case, $|^{57}A_{\rm hf}(\pi/a, \pi/a)|$ becomes largest, whereas $|^{75}A_{\rm hf}^\perp(\pi/a,\pi/a)|$ is expected to be zero. Thus $^{57}(1/T_{1}T)/^{75}(1/T_{1}T)$ would be the extremely large value, which is inconsistent with the experiment.

Thus far it has been theoretically proposed that the multiple spin-fluctuation modes with ${\bm Q}=$($\pi/a$, 0) and (0, $\pi/a$) arising from the nesting across the disconnected Fermi surfaces would mediate the extended s$_{\pm}$-wave pairing with a sign reversal of the order parameter\cite{Kuroki}. 
However, in our simple analyses, we could state only that the spin fluctuations at finite wave vectors is more significant than the ferromagnetic spin fluctuations mode in this compound. 
Nevertheless, it is noteworthy that the $(1/T_1T)$s for both the Fe and As sites decrease upon cooling, indicating the decrease of low-energy spectral weight of spin fluctuations over whole ${\bm q}$ space from the room temperature. 
In contrast, in the case of the copper-oxide superconductors, $1/T_{1}T$s of $^{63}$Cu and $^{17}$O exhibit a different $T$ dependence, due to the difference in the $q$-dependence of $^{63,17}A_{\rm hf}({\bf q})$ and strong AFM spin fluctuations with ${\bm Q}=(\pi/a,\pi/a)$.\cite{Takigawa}  
The suppression of spin fluctuations over whole ${\bm q}$ space upon cooling below room temperature was observed in FeAs-derived high $T_c$ superconductor, which has never been observed for other strongly correlated superconductors where an AFM interaction plays vital role in mediating the Cooper pairing. 
In this context, it is an open question at the present what type of fluctuations are responsible for a pairing glue taking place a possible extended s$_{\pm}$-wave pairing. 

In summary, the first $^{57}$Fe-NMR studies have unraveled the novel SC and normal-state characteristics for the  $^{57}$Fe-enriched LaFeAsO$_{0.7}$ with $T_{c}$ = 28 K. 
The measurements of the Knight shift and the $T_1$ of $^{57}$Fe have revealed that the extended s$_{\pm}$-wave pairing with a sign reversal of the order parameter can be a promising candidate in LaFeAsO$_{0.7}$ with $T_{c}$ = 28 K.
This novel SC is theoretically proposed to be realized by the multiple spin-fluctuation modes with $Q=$($\pi/a$, 0) and (0, $\pi/a$) arising from the nesting across the disconnected Fermi surfaces. 
However, in the normal state, we found the remarkable decrease of $1/T_1T$ upon cooling for both the Fe and As sites, indicating the decrease of low-energy spectral weight of spin fluctuations over whole ${\bm q}$ space upon cooling below room temperature, which has never been observed for other strongly correlated superconductors where an AFM interaction plays vital role in mediating the Cooper pairing.
Further experiments on $T_c$-dependences of $1/T_1$ and $K$ at both Fe and As sites by using a single crystal are required to understand the nature of spin fluctuations of this compound.


This work was supported by Grant-in-Aid for Specially promoted Research (20001004)  and in part by Global COE Program (Core Research and Engineering of Advanced Materials-Interdisciplinary Education Center for Materials Science), from the Ministry of Education, Culture, Sports, Science and Technology (MEXT), Japan.



\begin{thebibliography}{99} 
\bibitem{Kamihara} Y. Kamihara, T. Watanabe, M. Hirano, and H. Hosono: J. Am. Chem. Soc. {\bf 130} (2008) 3296.
\bibitem{Ren1} Z. A. Ren, W. Lu, J. Yang, W. Yi, X. L. Shen, Z. C. Li, G. C. Che, X. L. Dong, L. L. Sun, F. Zhou, and Z. X. Zhao: Chin. Phys. Lett. {\bf 25} (2008) 2215.
\bibitem{Kito} H. Kito, H. Eisaki, and A. Iyo: J. Phys. Soc. Jpn. {\bf 77} (2008) 063707.
\bibitem{Ren2} Z. A. Ren, G. C. Che, X. L. Dong, J. Yang, W. Lu, W. Yi, X. L. Shen, Z. C. Li, L. L. Sun, F. Zhou, and Z. X. Zhao: Europhys. Lett. {\bf 83} (2008) 17002.
\bibitem{C.H.Lee} C. H. Lee, H. Eisaki, H. Kito, M. T. Fernandez-Diaz, T. Ito, K. Kihou, H. Matsushita, M. Braden, and K. Yamada: J. Phys. Soc. Jpn. {\bf 77} (2008) 083704.
\bibitem{Mukuda} H. Mukuda, N. Terasaki, H. Kinouchi, M. Yashima, Y. Kitaoka, S. Suzuki, S. Miyasaka, S. Tajima, K. Miyazawa, P. M. Shirage, H. Kito, H. Eisaki, and A. Iyo: J. Phys. Soc. Jpn. {\bf 77} (2008) 093704.
\bibitem{FreemanWatson} A.J.Freeman and R.E. Watson, {\it in Hyperfine Interactions in Magnetic Materials, Magnetism IIA}, edited by G.T. Rado and S. Suhl ~Academic, New York, 1965.
\bibitem{Nakai} Y. Nakai, K. Ishida, Y. Kamihara, M. Hirano, and H. Hosono: J. Phys. Soc. Jpn. {\bf 77} (2008) 073701.
\bibitem{Grafe}	H.-J. Grafe, D. Paar, G. Lang, N. J. Curro, G. Behr, J. Werner, J. Hamann-Berrero, C. Hess, N. Leps, R. Klingeler, and B. B$\rm{\ddot u}$chner: Phys. Rev. Lett. {\bf 101} (2008) 047003.
\bibitem{Matano}  K. Matano, Z. A. Ren, X. L. Dong, L. L. Sun, Z. X. Zhao, and G.-q. Zheng: Europhys. Lett. {\bf 83} (2008) 57001.
\bibitem{Ahilan} K. Ahilan, F. L. Ning, T. Imai, A. S. Sefat, R. Jin, M. A. McGuire, B. C. Sales, and D. Mandrus: Phys. Rev. B {\bf 78} (2008) 100501(R).
\bibitem{ImaiJPSJ} T. Imai, K. Ahilan, F.L. Ning, M. A. McGuire, A. S. Sefat, R. Jin, B. C. Sales, D. Mandrus, J. Phys. Soc. Jpn. Suppl.(in press)
\bibitem{ARPES} H. Ding, P. Richard, K. Nakayama, K. Sugawara, T. Arakane, Y. Sekiba, A. Takayama, S. Souma, T. Sato, T. Takahashi, Z. Wang, X. Dai, Z. Fang, G. F. Chen, J. L. Luo and N. L. Wang: Europhys. Lett. {\bf 83} (2008) 47001. 
\bibitem{Penetrationdepth} K. Hashimoto, T. Shibauchi, T. Kato, K. Ikada, R. Okazaki, H. Shishido,
M. Ishikado, H. Kito, A. Iyo, H. Eisaki, S. Shamoto, and Y. Matsuda: arXiv:0806.3149.
\bibitem{Mazin} I.I. Mazin, D. J. Singh, M. D. Johannes, and M. H. Du: Phys. Rev. Lett. {\bf 101} (2008) 057003.
\bibitem{Kuroki} K. Kuroki, S. Onari, R. Arita, H. Usui, Y. Tanaka, H. Kontani, and H. Aoki: Phys. Rev. Lett. {\bf 101} (2008) 087004.
\bibitem{NMRtheory} A. V. Chubukov, D. Efremov, and I. Eremin: Phys. Rev. B {\bf 78} (2008) 134512; D. Parker, O.V. Dolgov, M.M. Korshunov, A.A. Golubov, and I.I. Mazin: Phys. Rev. B {\bf 78} (2008) 134524 ; Y. Bang, H.-Y. Choi: Phys. Rev. B {\bf 78} (2008) 134523; M. M. Parish, J. Hu, and B. A. Bernevig:arXiv:0807.4572. 
\bibitem{NMRtheory2} Y. Nagai, N. Hayashi, N. Nakai, H. Nakamura, M. Okumura, and M. Machida: arXiv:0809.1197.
\bibitem{Kitagawa} K. Kitagawa, N. Katayama, K. Ohgushi, M. Yoshida, and M. Takigawa: to be published in J. Phys. Soc. Jpn.
\bibitem{Takigawa} M. Takigawa, A. P. Reyes, P. C. Hammel, J. D. Thompson, R. H. Heffner, Z. Fisk, and K. C. Ott: Phys. Rev. B {\bf 43} (1991) 247. 
\end{thebibliography}
\end{document}